\documentstyle[twocolumn,eqsecnum,aps,prd]{revtex}
\begin{document}
%
%
%
%
\input boxedeps 
\HideDisplacementBoxes 
\SetUnixCoopEPSFSpecial
%
\def\title#1{\begin{center}{\Large #1}\end{center}}
\def\author#1{\vspace*{0.5ex}\begin{center}{#1}\end{center}}
\def\address#1{\begin{center}\vspace*{-1ex}{\it #1}\end{center}}  
\def\pubnum{330/COSMO-72}
\def\mark#1{\vspace*{1cm} \fbox{ #1}}
\def\NOTE#1{{\bf (*#1*)}}
\def\Isom#1{{\mbox{{\rm Isom}$#1$}}}
\def\UC#1{\widetilde #1}
\def\ME#1{#1'}
\begin{titlepage}
  \hfill
  \parbox{6cm}{{TIT/HEP-\pubnum} \par May, 1996 }
  \par
  \vspace{21mm}
  \title{\bf Compact hyperbolic universe and singularities}
  \vspace{5mm}
  \author{Akihiro ISHIBASHI\footnote{E-mail: akihiro@th.phys.titech.ac.jp}} 
  \address{Department of Physics,\\  Tokyo Institute of Technology,
    Oh-Okayama, Meguro, Tokyo 152, Japan}
  \author{Tatsuhiko KOIKE\footnote{E-mail: koike@rk.phys.keio.ac.jp}} 
  \address{Department of Physics,\\  Keio University,
    Hiyoshi, Kohoku, Yokohama 223, Japan}
  \author{Masaru SIINO\footnote{E-mail: msiino@tap.scphys.kyoto-u.ac.jp, 
      JSPS fellow}}
  \address{Department of Physics,\\ Kyoto University,
    Kyoto 606-01, Japan}
  \author{Sadayoshi KOJIMA\footnote{E-mail: sadayosi@is.titech.ac.jp}}
  \address{Department of Mathematical and Computing Science, \\
    Tokyo Institute of Technology,
    Oh-Okayama, Meguro, Tokyo 152, Japan}
  \begin{abstract}
    \large
    \begin{center}
      {\bf Abstract}
    \end{center}
    Recently many people have discussed the possibility 
    that the universe is hyperbolic and was in an inflationary phase 
    in the early stage.
    Under these assumptions, it is shown that the universe cannot have 
    compact hyperbolic time-slices.  
    Though the universal covering space of the universe has 
    a past Cauchy horizon and can be extended analytically beyond it,
    the extended region has densely many points which correspond to 
    singularities of the compact universe.
    The result is essentially attributed 
    to the ergodicity of the geodesic flow on a compact negatively curved 
    manifold.   
    Validity of the result is also discussed in the case of
    inhomogeneous universe. 
    Relationship with {\it the strong cosmic censorship 
    conjecture}\/ is also discussed. 
  \end{abstract}
\end{titlepage}

\section{Introduction}


Recent observations of the density parameter 
$\Omega_0\approx 0.1$~\cite{Fuku} 
suggest that the spatial curvature of 
the universe is negative. In this paper we call such the universe 
with negative curvature as a hyperbolic universe rather than a 
conventional open universe because the {\it open}\/ universe is meant 
to be non-compact while we shall consider a {\it compact}\/ negative 
curvature space.     


On the other hand, it is believed that inflations occurred in 
the early stage of the universe because the Inflationary scenario is, 
so far, the only viable theory which can solve such cosmological problems 
as the flatness, horizon and monopole problems 
\cite{INF} without appealing to fine tuning. 
It is therefore of great interest to investigate 
hyperbolic inflationary universe scenarios. 

There have been some works which studied the creation of the 
hyperbolic inflationary universe in the one-bubble inflationary 
model~\cite{OBOU}.  


During the inflation epoch, the universe is well approximated by a de Sitter 
space-time. Then, we notice that hyperbolic universe has no 
curvature singularity as long as it is in the inflationary phase, 
even though its spatial volume approaches to zero. 
The initial singularity is just a coordinate singularity. 
Hyperbolic inflationary universe is expected to be realized 
by taking hyperbolic time-slices on a portion of the de Sitter space-time. 
One example is the one-bubble inflationary scenario. 
In this context, any space-time which contains a hyperbolic 
inflationary universe has a past (Cauchy) horizon and continues further 
to the past because a hyperbolic chart 
cannot cover the whole de Sitter space-time. 


Observations normally give us local information only such as 
the local spatial curvature of our universe, because local information is, 
in general, independent of global properties such as the topology of 
the universe. Some observational effects are, however, expected, 
if the periodicities due to the non-trivial topology are shorter than 
the horizon scale. Such possibilities have been studied in theoretical 
and observational cosmologies~\cite{CU}. 

Compactness of the universe seems to be an acceptable 
and appealing notion, especially in the context of 
the canonical treatments of the universe (or quantum gravity). 
Compactness provides a finite value of the action integral 
and gives the natural boundary conditions for the matter and gravitational 
fields in the universe. 

The notion of compactness of a given universe depends on how to 
take its spatial sections. We concentrate our interests 
on the case that hyperbolic hypersurfaces are compact. 
In particular, an intriguing question arises in the case that 
compact hyperbolic universe is in an inflationary phase. 
We naturally ask whether or not such universe has the past 
horizon and continues further to the past as the non-compact case. 
If it does, we expect that topology change takes place and closed timelike 
curves appear, which is suggested by the work on the Lorentzian
topology change of (2+1)-dimensional compact black hole geometry 
by M. Siino~\cite{Siino}. 
If it does not, we may find restriction to possible topologies of the universe. 


In this paper, we will study the spatially compact hyperbolic 
inflationary universe. We will find that the latter is indeed 
the case. Namely, we will show that compact hyperbolic 
universe cannot be extended beyond the past horizon. 
This means that if the universe is hyperbolic and was in an inflationary phase 
in the early stage, the universe is not spatially compact. 


In the next section, we construct models of compact hyperbolic 
inflationary universe. In section~3, we examine the extendibility of the universe. 
Section~4 is for conclusions and discussions.

\section{Compact hyperbolic inflationary universe}

\subsection{Construction of compact universe}


In this section, we treat homogeneous isotropic models of hyperbolic 
universe in an inflationary phase. Such a model is described 
by a hyperbolic chart on 4-dimensional de Sitter space-time. 
In the hyperbolic chart, homogeneous isotropic spacelike hypersurface 
is a manifold $H^3$ of constant negative curvature with isometry group $SO(3,1)$. 
This universe has a past Cauchy horizon, and in the extended region 
a homogeneous hypersurface becomes 3-dimensional de Sitter hypersurface $dS^3$, 
which is an orbit of $SO(3,1)$. 


We construct the compact universe model by identifying points of the space-time. 
A compact hyperbolic hypersurface is given as 
the quotient space of $H^3$ by the 
discrete subgroup $\Gamma$ of its isometry group $SO(3,1)$. 
One of the simplest 3-dimensional compact 
hyperbolic manifold is known as the Seifert--Weber manifold~\cite{WS}, 
whose construction is explicitly shown in appendix A. 

 
In general, the action of a group $\Gamma$ on a manifold 
$M$ must be properly discontinuous so that the 
quotient space $M/\Gamma$ be a Hausdorff manifold. 
We call an action of $\Gamma$ on $M$ {\it properly
discontinuous} if it satisfies the following conditions~
\cite{HE,KN}; 
\begin{enumerate}
\item each point $q\in M$ has a neighborhood $U$ such that 
  $\gamma(U)\cap U = \emptyset$ for each $\gamma\in \Gamma$ 
  which is not the identity element, and  
\item if $q,r \in M $ are such that there is no $\gamma \in \Gamma$ 
  with $\gamma(q)=r$, then there are neighborhoods $U$ and $U'$ 
  of $q$ and $r$ respectively such that there is no 
  $\bar{\gamma}\in \Gamma$ 
  with $\bar{\gamma}(U) \cap U' \neq \emptyset$. \\ 
\end{enumerate}
Condition~(1) implies that the quotient $M/\Gamma$ is a 
manifold, and condition~(2) implies that it is Hausdorff. 

It is worth noting that, for a Riemannian manifold 
$(M,g)$, every {\it discontinuous} group $\Gamma$ of Isom$(M,g)$ is 
properly discontinuous~\cite{KN}. 
Here the action of $\Gamma$ on a Riemannian manifold is 
called discontinuous if for every $p \in M$ and every sequence $\{\gamma_i\}$ of 
elements of $\Gamma$, where all $\gamma_i$ are 
mutually distinct, the sequence $\{ \gamma_i (p)\}$ 
does not converge to a point in $M$~\cite{KN}. 
We can obtain the compact hyperbolic inflationary universe 
because the homogeneous hypersurface $H^3$ is a Riemannian 
manifold. 

\subsection{Misner universe}

As the simplest example of a compact universe and 
its extension, let us see the construction of Misner 
universe, which is given as a quotient manifold
in (1+1)-dimensions.

The Misner universe $(M, g_M)$ is a space-time diffeomorphic to 
$S^1 \times {\bf R}$. Its metric is given by 
\begin{equation}
g_M = -t^{-1}dt^2 + t d\psi^2,\quad t\in {\bf R},\quad 
0 \le \psi \le 2\pi . 
\end{equation} 
The universal covering manifold is the region $(I,\widetilde{\eta})$ in Figure 1, 
which is a part of 2-dimensional Minkowski space-time 
$(\widetilde{M},\widetilde{\eta})$. The isometry group of 
$(\widetilde{M},\widetilde{\eta})$ is the Lorentz group $SO(1,1)$, 
whose orbits are the hyperbolae $(X^0)^2-(X^1)^2 =:\sigma= \mbox{constant}$. 
The covering transformation group $\Gamma$ 
of the Misner universe is a discrete subgroup 
of the Lorentz group 
consisting of $\gamma^m$, where $m$ is an integer and $\gamma$ 
maps $(X^0, X^1)\in \widetilde{M}$ to 
\begin{equation}
    \pmatrix{{X^0}' \cr {X^1}'} = 
\pmatrix{\cosh\pi & \sinh\pi \cr
        \sinh\pi & \cosh\pi \cr } \pmatrix{X^0 \cr X^1}. 
\end{equation}
This subgroup $\Gamma$ acts on the region 
$(I,\widetilde\eta)$ properly discontinuously. 
However, if we maximally extend the Misner universe, for any point $r$ on 
the null line $X^0 + X^1 = 0$, the sequence $\{\gamma^m (r)\}$ has an 
accumulation point $o:=(0,0)$ on $(\widetilde{M},\widetilde{\eta})$. 
Hence the condition (1) of properly discontinuous is not satisfied and the quotient 
$\widetilde{M} /\Gamma$ fails to be a manifold.
We therefore must remove the accumulation point $o$ 
from the whole Minkowski space-time $(\widetilde{M},\widetilde{\eta})$ 
so that we obtain the maximally extended Misner universe 
as a manifold $(\widetilde{M} \setminus \, \{o\})/\Gamma$, 
which is illustrated in Figure 1. 
Since the action of $\Gamma$ on $\widetilde{M} \setminus \, \{o\}$ 
does not satisfy the condition (2), 
$(\widetilde{M} \setminus \, \{o\})/\Gamma$ is a non-Hausdorff manifold.   

Because of removing the point $o$ from 
$(\widetilde{M}, \widetilde{\eta})$, 
all causal curves toward $o$ are incomplete. 
We define a singularity as an incomplete causal curve 
according to Hawking and Ellis~\cite{HE}. 
Thus, the maximally extended Misner universe has a singularity. 
This indicates that the existence of the accumulation points of action of 
$\Gamma$ causes singularities of the quotient space-time. 
In this paper, we call an accumulation 
point {\it a topological singularity}. The null boundaries of the original 
region $I$ and the extended regions $I\!I$, $I\!I\!I$ are compact 
Cauchy horizons, which are generated by closed null geodesics. 
\begin{figure}
\begin{center}
{\BoxedEPSF{Misner.eps scaled 340} }
\end{center}
 \caption{The maximally extended Misner universe is illustrated. 
    The coordinate origin $o$ is an accumulation point of a sequence 
    $\{ \gamma^m (r) \}$. The region $I$ is the original Misner universe region. 
    Under the action of $\Gamma$, points $s$ on the $\sigma$-constant surface 
    are equivalent; similar for points $r$, points $q$, and points $t$, respectively.
    In the extended regions $I\!I$, $I\!I\!I$, closed timelike curves appear.}
\protect \label{fig:Misner}
\end{figure}
\section{Anomalous occurrence of singularities}
 
In case of the compact hyperbolic inflationary universe model, 
the discrete subgroup $\Gamma$ of $SO(3,1)$ does not act on 
hypersurface $dS^3$ in the extended region properly discontinuously. 
Moreover, in contrast with the Misner universe, we will observe that 
topological singularities densely appear on the hypersurfaces $dS^3$ and 
the extended region is no longer a space-time manifold. 
Our main result is the following proposition.
\begin{quote}
  {\bf Proposition.} \\
  {\it Suppose there exists an analytic extension of the universal cover 
    of $(n+1)$-dimensional compact hyperbolic universe, where $n\ge2$.
    Then, for every neighborhood $O_c$ 
    of an arbitrary point $c$ in the $dS^n$, 
    there exist $\gamma \in \Gamma$ and point $s\in dS^n$ 
    such that an accumulation
    point $c'$ of the infinite sequence $\{ \gamma^m (s)\}$ 
    is contained in $O_c$.} 
\end{quote}


Under the assumption of the analyticity, we naturally observe 
that there exist a neighborhood which contains a Cauchy horizon in 
the hyperbolic inflationary universe and that in a Milne universe which 
are analytically diffeomorphic to each other. 
Here the Milne universe is the Minkowski space-time $E^{n,1}$ 
with the hyperbolic chart. Thus it is convenient to exam in the Milne universe 
instead of the hyperbolic inflationary universe. 

Since the argument is the same for $(n+1)$-dimensional universe with
$n\ge2$, we first present our argument in the $n=2$ case and comment
on the $n\ne3$ cases later.


Let us investigate what the action of $\Gamma$ is like 
in the extended region of the $(2+1)$-dimensional compact hyperbolic
inflationary universe. As mentioned above, this is done 
in the 3-dimensional Milne universe.
Let $E^{2,1}$ be a $3$-dimensional Minkowski space-time with 
the metric 
\begin{equation}
  \eta =-(dX^0)^2+(dX^1)^2+(dX^2)^2. 
\end{equation} 
In the model the hyperbolic hypersurface is 2-hyperbolic space $H^2$, 
which is embedded in $E^{2,1}$ as 
\begin{equation}
 -(X^0)^2+(X^1)^2+(X^2)^2 = -l^{2},\label{eq:H2} 
\end{equation} 
by embedding $f: H^2 \longrightarrow E^{2,1}$, 
where $l$ is the curvature radius. 
The induced metric is given by $g_H := f^{\ast}\eta$ and whose 
isometry group is $SO(2,1)$.   

In the extended region of the Milne universe the orbit of $SO(2,1)$ is $dS^2$, 
which is a surface  
\begin{equation}
 -(X^0)^2+(X^1)^2+(X^2)^2 = l^{2},\label{eq:dS2}
\end{equation}
in $E^{2,1}$. 
Each element of the subgroup $\Gamma$ must contain one of 
the boosts because otherwise the element consisted 
of only space-rotations would have fixed points. 
We consider the action of the element $\gamma_{\circ} \in \Gamma$ on 
$dS^2$, where $\gamma_{\circ}$ is represented in the coordinate system 
$(X^0, X^1, X^2)$ by the following matrix; 
\begin{equation} 
\gamma_{\circ}\left( \zeta \right) := \pmatrix{
      \cosh\zeta & 0 & \sinh\zeta \cr
         0 & 1 & 0 \cr
      \sinh\zeta & 0 & \cosh\zeta \cr
} ,\quad  \zeta = {\rm constant}. 
\end{equation}
$\gamma_{\circ}$ has three eigenvectors, 
\begin{equation}
{a_{\circ}} := \left( \begin{array}{c}
1\\0\\-1 \end{array} \right), \quad 
{b_{\circ}} := \left( \begin{array}{c}
1\\0\\1 \end{array} \right), \quad
{c_{\circ}} := l\left( \begin{array}{c}
0\\1\\0 \end{array} \right). 
\end{equation}
The two eigenvectors ${a_{\circ}}$ and 
${b_{\circ}}$ are lying on the light cone, 
\begin{equation}
-(X^0)^2 +(X^1)^2 +(X^2)^2 = 0, \label{eq:nullcone}
\end{equation}
and the point $c_{\circ}$ is in the $dS^2$ described 
by Eq.~(\ref{eq:dS2}). Hereafter we use the same symbol to denote 
eigenvectors and the endpoints of the arrows of the eigenvectors. 
As in the maximally extended Misner universe, we observe that any point $s$ 
on the null lines in the $dS^2$ through the point $c_{\circ}$ converges to 
$c_{\circ}$ by the action of ${\gamma_{\circ}}^m$ as illustrated in 
Figure~\ref{fig:eigenflow}. That is, the point $c_{\circ}$ is the accumulation point 
of the infinite sequence $\{{\gamma_{\circ}}^m(s)\}$. 
For any $\gamma (\neq \gamma_{\circ}) \in \Gamma$, 
$\gamma({c_{\circ}})$ is one of the eigenvectors of 
${\gamma} \circ \gamma_{\circ} \circ {\gamma}^{-1}
\in \Gamma$ and so the point $\gamma({c_{\circ}})$ in the $dS^2$ is also 
the accumulation point of the sequence 
$\{({\gamma \circ \gamma_{\circ} \circ {\gamma}^{-1}})^m(\gamma(s))\}$. 
Thus, there exist many, countably infinite, accumulation points 
of the sequences $\{{\gamma}^m(s)\}$ in the $dS^2$, if we consider all
$\gamma\in\Gamma$.   
\begin{figure}
\begin{center}
{\BoxedEPSF{eigenflow.eps scaled 340} }
\end{center}
 \caption{Orbits of ${\gamma_{\circ}}^m(p), \, p\in dS^2$ 
  are schematically depicted. }
 \protect \label{fig:eigenflow}
\end{figure}

Now we demonstrate that the accumulation points, that are topological 
singularities, densely occur in $dS^2$. 

First, for every point $c$ in the $dS^2$, there exist an element 
$\gamma_c \in SO(2,1)$ such that ${c}$ is one of the eigenvectors 
of $\gamma_c$. If $\gamma_c \in \Gamma$, there exists $s\in dS^2$ 
such that $c$ is the accumulation point of $\{ \gamma_c^m (s)\}$. 

In the case that $\gamma_c \notin \Gamma $, the other two eigenvectors, 
${a}$ and ${b}$, of $\gamma_c$ are lying on the light cone (Eq.(\ref{eq:nullcone})). 
These two vectors uniquely determine a geodesic 
in the $H^2$, as depicted in Figure \ref{fig:Proj}. 
In fact, these two eigenvectors span 
a plane $\Pi$ and the intersection $\Pi \cap H^2$ is an 
geodesic curve $\lambda $ in the $H^2$. 
The geodesic $\lambda$ is projected to a straight line 
$\lambda_K:=\Pi \cap D_K$ on {\it the Klein Disk}
$D_K := \{ (X^0,X^1,X^2)|X^0 = 1,\, (X^1)^2 + (X^2)^2 < 1 \}$. 
The projection $\pi : E^{2,1}\ni (X^0,X^1,X^2) \mapsto 
(k_1, k_2) \in D_K$ is defined as 
\begin{equation}
 k_i := \frac{X^i}{X^0} \quad (i=1,2). 
\end{equation}
The induced metric takes the form
\begin{eqnarray}
  g_K = \frac{1}{(1-k^2)^2}
  \left [ (1-{k_2}^2)d{k_1}^2 + 
  (1-{k_1}^2)d{k_2}^2 \right.
  \nonumber\\
  \left. + 2 k_1 k_2 d{k_1}d{k_2} \right ],
\end{eqnarray}
where $k := \sqrt{{k_1}^2+{k_2}^2}$.
We can identify the $H^2$ and $D_K$ by the diffeomorphism $\pi \circ f$. 
This projection $\pi$ maps a geodesic curve of the $H^2$, which is
a hyperbola in the $E^{2,1}$ to a straight line in the $E^{2,1}$.
We can also identify the action of $\Gamma$ on the $H^2$ and the action of
$\Gamma_K$ on $D_K$, where 
\begin{equation}
  \gamma_K := \pi\circ\gamma\circ\pi^{-1},\quad
  \Gamma_K := \{\gamma_K | \gamma\in\Gamma\}.
\end{equation}
\begin{figure}
\begin{center}
{\BoxedEPSF{Proj.eps scaled 340} }
\end{center}
        \caption{Projection $\pi:E^{2,1}\longrightarrow D_K$. The
          $H^2$ and $D_K$ is identified by $\pi\circ f$.  }
        \protect \label{fig:Proj}
\end{figure}

We can see that the straight line from $b$ to $a$ on 
$\overline{D}_K := D_K \cup \partial\!D_K$ 
is the geodesic $\lambda_K$, where 
$\partial D_K:= \{ (X^0,X^1,X^2)|X^0=1,(X^1)^2+(X^2)^2=1\}$.

As long as considering the case that $\gamma_{cK} \notin \Gamma_K$, 
the geodesic $\lambda_K$ is not closed on $D_K/\Gamma_K$. 

It can be observed that 
$\Xi:= \partial {D_K} \times \partial {D_K} \setminus 
\Delta_{\partial {D_K}}$ has one-to-one correspondence with 
the set of all geodesics on $D_K$, 
where $\Delta_{\partial {D_K}}:= \{(p,p)|\,p \in \partial D_K \}$. 
Namely, a pair $(a,b) \in \Xi$ can be identified with the geodesic 
on $D_K$ which has endpoints $a, b\in \partial D_K$. 

The eigenvectors $a$, $b$ and $c$ 
are related to each other as follows. Choosing an appropriate 
coordinate system and a constant $\alpha$, 
the eigenvector ${c}$ can be represented as 
\begin{equation}
{c} = l\left( \begin{array}{c}
\sinh\alpha \\ \cosh\alpha \\0 \end{array} \right).    
\end{equation}
Then, the other two are uniquely determined as 
\begin{equation}
{a} = \left( \begin{array}{c}
\cosh\alpha \\ \sinh\alpha \\-1 \end{array} \right), \quad 
{b} = \left( \begin{array}{c}
\cosh\alpha \\ \sinh\alpha \\1 \end{array} \right),  
\end{equation} 
except for the freedoms of the norms. 
Accordingly, for any point $c\in dS^2$, a geodesic $\lambda$ on the $H^2$ 
(or a point in $\Xi$) is uniquely determined.\footnote{If the point $c$ is 
constrained on the part $\{(X^0,X^1,X^2) |$ $X^0 >0\}$ $\cup$
$\{(X^0,X^1,X^2)|X^0 = 0$, $(X^1)^2 + (X^2)^2 =1$, $X^1 >0 \}$
$\cup$ $\{(0,0,1)\}$ of the $dS^2$, the point $c$ has one-to-one 
correspondence with a point in $\Xi$.} 

Second, choose a Riemannian metric ${\bf e}:=(dX^0)^2+(dX^1)^2+(dX^2)^2$ 
on $E^{2,1}$. 
Take an open ball, $B_c(\delta):= \{x\in E^{2,1}|\,\|x-c\|_{\bf e} <\delta\}$, 
of radius $\delta$ with respect to ${\bf e}$ and define 
a neighborhood of $c$ in the $dS^2$ as 
$O_c(\delta, dS^2):= B_c(\delta)\cap dS^2$. 
Similarly, define neighborhoods of $a$ and $b$ respectively 
in the $\partial D_K$ as $O_a(\delta, \partial D_K):= 
B_a(\delta)\cap \partial D_K$ and 
$O_b(\delta, \partial D_K):= 
B_b(\delta)\cap \partial D_K$ 
corresponding to $O_c(\delta, dS^2)$ (see Figure \ref{fig:delta}). 
Then, for any point $c'\in O_c(\delta, dS^2)$, there exist 
$a' \in O_a(\delta, \partial D_K)$, $b' \in O_b(\delta, \partial D_K)$ 
and ${\gamma_{c'}} \in SO(2,1)$ such that $a'$, $b'$ and $c'$ are 
the eigenvectors of ${\gamma_{c'}}$.
  
\begin{figure}
\begin{center}
{\BoxedEPSF{delta.eps scaled 340} }
\end{center}
        \caption{Neighborhoods of points $a, b\in \partial D_K$ and 
        $c\in dS^2$.}
        \protect \label{fig:delta}
\end{figure}

Now, what we want to show is reduced to the following lemma; \\
\begin{quote}
  {\bf Lemma.} \\
  {\em For any $(a, b) \in \Xi$ and $\delta > 0$, 
    there exist $a'\in O_a(\delta, \partial D_K)$,
    $b'\in O_b(\delta, \partial D_K)$, 
    $\gamma_K \in \Gamma_K$ and 
    $p \in D_K $ such that 
    \begin{eqnarray*}
      \lim_{n \to \infty}\gamma_K^m (p)
      &=& a' \in O_a(\delta, \partial D_K)\\ 
      \lim_{n \to -\infty}\gamma_K^m (p) 
      &=& b'\in O_b(\delta, \partial D_K).
    \end{eqnarray*}
    }
\end{quote}

\noindent{\em Proof.}\\
We show that on the compact hyperbolic manifold $D_K /\Gamma_K$ 
there exists a closed curve
whose lift has endpoints $a'\in O_a(\delta, \partial D_K)$, 
$b'\in O_b(\delta, \partial D_K)$ in $\partial D_K$. 
We can make such a closed curve by using a non-closed geodesic 
$\lambda_K (v) := \Pi \cap D_K$ affinely parameterized by $v$ 
and ergodicity of geodesic flow on a compact manifold 
with negative curvature (see appendix B). 
\begin{figure}
 \begin{center}
{\BoxedEPSF{pqr.eps scaled 340} }
\end{center}
 \caption{A compact hyperbolic manifold $D_K /\Gamma_K$ with genus 2
   is illustrated. $\lambda_K$ is a non-closed geodesic 
   and $\mu$ is a line segment which connects two points $q$ and $r$ 
   on the $\lambda_K$. Then, one can see 
   a closed curve
   $(r \rightarrow p \rightarrow q \rightarrow r )$ 
   composed of the line segment 
   $( r\rightarrow p\rightarrow q)$ of the 
   geodesic $\lambda_K$ and the line segment $\mu$ 
   $(q \rightarrow r)$. }
 \protect 
        \label{fig:pqr}
\end{figure}
Let $p$ be a point on $\lambda_K$ at 
$v=0$ and take an arbitrary small open neighborhood
$O^K_p(\epsilon, D_K/\Gamma_K) 
:= \{ x\in D_K/\Gamma_K | d_K(x,p)<\epsilon \}$
of $p$,
where $d_K$ is the distance naturally defined by $g_K$.
From the ergodicity of geodesic $\lambda_K$, 
for any large $N>0$, there exist $v_1>N,v_2<-N$ such that
$q:=\lambda_K(v_1)\in O^K_p$ and
$r:=\lambda_K(v_2)\in O^K_p$
and the tangent vectors of $\lambda_K$ at these 
points $r$, $p$, $q$, are sufficiently parallel to each other as 
depicted in Figure \ref{fig:pqr}. Connecting the points $r$ and 
$q$ by a suitable line segment $\mu$, we obtain a closed curve
$(r \rightarrow p \rightarrow q \rightarrow r )$ on the 
$D_K/\Gamma_K$. Corresponding to this closed curve, there exists 
an element $\gamma_K\in \Gamma_K$. 

\begin{figure}
 \begin{center}
   {\BoxedEPSF{Klein.eps scaled 340} }
  \end{center}
  \caption{A lift $\widetilde O^K_p$ and its images 
   $\gamma_K^m(\widetilde O^K_p)$ 
   on $\widetilde D_K$.}
  \protect \label{fig:Klein}
\end{figure}

Let us fix a component $\widetilde O^K_p$ of the lift of $O^K_p$ 
on $D_K$, which is diffeomorphic to $O^K_p$. Accordingly, 
$\tilde p$, $\tilde q$, $\tilde r$, $\lambda_K\cap \widetilde 
O^K_p$ and $\tilde \mu$ denote the corresponding lifts.
Any component of the lift of $O^K_p$ is given by 
$\gamma_K^m(\widetilde O^K_p)$.
These are illustrated in Figure \ref{fig:Klein}.
For any $N>0$ there exist $m_1, m_2>N$, $v_1>0$ and $v_2<0$ such that
$\lambda_K(v_1)\in\gamma_K^{m_1}(\widetilde O^K_p)$ and
$\lambda_K(v_2)\in\gamma_K^{m_2}(\widetilde O^K_p)$.

Let $g_E$ denote the Euclidean metric on $\overline D_K$
induced by ${\bf e}$. Then the radius of 
$\gamma_K^m(\widetilde O^K_p)$ measured by $g_E$ becomes 
smaller and smaller as $m\rightarrow\pm\infty$. 
We therefore have 
$\lim_{m\rightarrow\infty}\gamma_K^m(\widetilde O^K_p)
\subset O_a(\delta,\partial D_K)$ and 
$\lim_{m\rightarrow -\infty}\gamma_K^m(\widetilde O^K_p)
\subset O_b(\delta,\partial D_K)$.
As a result, we obtain
\begin{eqnarray*}
  &&a':= \lim_{m \to \infty}\gamma_K^m (p) \in O_a(\delta, 
\partial D_K), \\
  &&b':= \lim_{m \to -\infty}\gamma_K^m (p)\in O_b(\delta, 
\partial D_K). 
\end{eqnarray*}
\hfill$\Box$

{\em Proof of the Proposition.} \\ 
From the Lemma we immediately 
have the two eigenvectors ${a'}$ and ${b'}$ of $\gamma\in\Gamma$.
Then we obtain the third eigenvector ${c'}$ of $\gamma$ 
so that the point $c'$, which is the accumulation point of
$\{\gamma^m(s)\}$,  is contained in $O_c(\delta,dS^2)$. 
This is the proof of $n=2$ case.


In $(3+1)$-dimensional case, the discrete subgroup of $SO(3,1)\cong{\rm Isom}(H^3)$ 
has four eigenvectors. Two of them are null vectors 
corresponding to ${a}$ and ${b}$ in the $(2+1)$ case. 
The other two corresponding to ${c}$ are spacelike and direct to 
points in the hypersurface $dS^3$. In addition, 
we can also observe the ergodicity of geodesic flows on three compact 
hyperbolic Riemannian manifold $H^3/\Gamma$~\cite{AA}(see appendix B). 
We immediately obtain the same result as the $(2+1)$-dimensional 
case; topological singularities densely occur in the $dS^3$. 

The assertion is proven in a similar manner in the higher dimensional cases.
\hfill$\Box$

\section{Conclusions and discussion}


We have demonstrated that if a space-time with spatially compact 
hyperbolic hypersurfaces is extended analytically 
by extending both the universal cover and the 
action of $\Gamma \subset $ Isom($M$) thereof, the topological 
singularities appear densely in the extended region and it is no longer 
a manifold. We conclude that the spatially compact hyperbolic inflationary 
universe cannot be extended beyond the Cauchy horizon of its 
universal cover. It follows that the universe cannot be compact, 
if the universe is hyperbolic and was in an inflationary phase. 
From observations, it is difficult to determine global properties of 
the universe. It is full of interest that our result theoretically 
answers to some global properties, for example, to the simple but fascinating 
question whether our universe is spreading infinitely or compact. 

Our result that the universe cannot be extended analytically 
across the Cauchy horizon
 obviously holds in the case that 
the spatial sections are ${\bf R} \times H^2/\Gamma$, which is non-compact,
or its quotient manifold $S^1 \times H^2/\Gamma$. 
It is remarkable that this is true even if the metric is
{\em inhomogeneous}\/ along the ${\bf R}$ (or $S^{1}$) factor.
Consequently, the universe with such a topology does not have 
a Cauchy horizon. 


As the scenario for a birth of hyperbolic universe, 
one-bubble inflationary universe scenario is an appealing one. 
However, it cannot be a universal covering manifold of a compact hyperbolic 
inflationary universe, because such a model has a past 
Cauchy horizon inside the bubble. Thus, one-bubble
inflationary universe scenario is incompatible 
with a compact hyperbolic universe model. 
In other words, if the one-bubble inflationary scenario is verified by observations, 
our hyperbolic Friedmann--Robertson--Walker (FRW) universe is not compact.
In general, any scenario which realize the hyperbolic inflationary universe 
by inducing hyperbolic chart on a portion of de Sitter space-time has 
a Cauchy horizon for the hyperbolic hypersurfaces 
and hence it cannot be a universal cover of a compact 
hyperbolic inflationary universe. 

As we show explicitly, our result holds for the (2+1)-dimensional 
case and anti-de Sitter space-time, compact 
3-dimensional black hole geometry does not realize 
Lorentzian topology change~\cite{Siino}. 


One may say that this anomaly is due to the high degrees of space-time 
symmetry. However we can also discuss the case that there is no 
symmetry. The result is essentially attributed to the ergodicity 
of geodesics on compact manifold with negative curvature~\cite{AA}. 
When we consider the inhomogeneous universe as a hyperbolic 
inflationary FRW universe model with perturbations on it, 
we can take a hypersurface whose sectional
curvature is everywhere negative by taking a time-slice of
sufficiently small scale factor, or $l$ in the Eq.~(\ref{eq:H2}) and 
Eq.~(\ref{eq:dS2}), near the past Cauchy horizon of the background universe. 
If inhomogeneity due to fluctuations of matter
fields is large enough, energy density of the matter fields dominates  
the universe and it is out of the context of the inflationary universe. 
In such a case, an initial curvature singularity appears instead of the past 
Cauchy horizon.  Thus, even in the case that the considering 
universe model is inhomogeneous, we expect that if the Cauchy horizon 
exists in its universal cover there exists a neighborhood 
of the Cauchy horizon such that the hypersurfaces contained in it 
are everywhere negatively curved. 
Then we observe that this neighborhood is 
homeomorphic to the neighborhood of the Cauchy horizon in 
the Milne universe, as considered in the previous section. 
We therefore expect the same result of the Proposition even in the inhomogeneous 
case. 

Our result that compact hyperbolic universe does not admit 
a Cauchy horizon is closely related to 
{\it the strong cosmic censorship conjecture}, 
which states that physically realistic space-time is globally 
hyperbolic~\cite{Penrose,ME}. 
The case of spatially compact, locally homogeneous vacuum models 
have been extensively investigated by P. T. Chru\'sciel and A. D. Rendall~\cite{CR}. 
Our result is not restricted to the vacuum case 
and our approach will be useful to resolve the issue in the case of 
inhomogeneous universe. 

\section*{Acknowledgments}
We are grateful to Prof. A.~Hosoya, Prof. H.~Ishihara and Prof. T.~Mishima 
for helpful discussions. 
T.~K. and M.~S. acknowledge financial supports from the Japan Society for the 
Promotion of Science and the Ministry of Education, Science and Culture.

\appendix
\section{Construction of three-compact hyperbolic manifold}

3-dimensional hyperbolic space $H^3$ can be embedded 
in 4-dimensional Minkowski space-time $E^{3,1}$ as  
$$
-(X^0)^2+(X^1)^2+(X^2)^2+(X^3)^2 = -1,
$$ 
where the curvature radius is normalized to 
unity. 
Taking the chart;
\begin{equation}
 \left\{
     \begin{array}{lllll}
      X^0 = \cosh \xi  ,\\
      X^1 = \sinh \xi \cos \theta  , \\
      X^2 = \sinh \xi \sin \theta \cos \psi ,\\
      X^3 = \sinh \xi \sin \theta \sin \psi ,         
     \end{array}
   \right. \label{eq:hcha} 
\end{equation} 
the induced metric takes the form  
\begin{equation}
  g_{H}=d\xi^2+\sinh^2\!\xi(d\theta^2 + 
\sin^2\! \theta d\psi^2).    
\end{equation}
Isom$(H^3)$ is $SO(3,1)$.

The simply connected Riemannian manifold $H^3$ can be 
compactified by quotienting by the subgroup $\Gamma$ of 
its isometry $SO(3,1)$. 
It is known that $H^3$ is tessellated by hyperbolic
dodecahedra. 
This means that the fundamental region of $H^3/ \Gamma$ is a 
hyperbolic dodecahedron~\cite{WS}.  
Then the concrete representations of the generators 
of $\Gamma$ are given 
as the following six matrices 
$\{  T_{i=1\sim 6}  \}$ under the coordinates 
$(X^0, X^1, X^2, X^3)$ of $E^{3,1}$;
\begin{eqnarray}
 T_1 &:=& B_{0-3}\left(\alpha \right) \circ R_3 \left
( \frac{3}{5}\pi \right)
 ,\\
 T_2 &:=& R_2 \left(2\psi \right) \circ T_1 \circ 
R_2\left( -2\psi \right) , \\
 T_3 &:=& R_3 \left( \frac{2}{5}\pi \right) \circ T_2 \circ 
R_3 \left( -\frac{2}{5}\pi \right) , \\
 T_4 &:=& R_3 \left( \frac{4}{5} \pi \right) \circ T_2 \circ 
R_3 \left( -\frac{4}{5}\pi \right)  ,  \\
 T_5 &:=& R_3 \left( -\frac{4}{5}\pi \right) \circ T_2 \circ 
R_3 \left( \frac{4}{5}\pi \right) , \\
 T_6 &:=& R_3 \left( -\frac{2}{5}\pi \right) \circ T_2 \circ 
R_3 \left( \frac{2}{5}\pi \right) ,
\end{eqnarray}
where
\begin{equation} B_{0-3}\left( \alpha \right) := \pmatrix{
      \cosh\alpha & 0 & 0 & \sinh\alpha \cr
         0 & 1 & 0 & 0 \cr
         0 & 0 & 1 & 0 \cr
      \sinh\alpha & 0 & 0 & \cosh\alpha \cr
} , 
\end{equation}
\begin{equation}
 R_2 \left( 2\psi \right) := \pmatrix{
      1 & 0 & 0 & 0 \cr
      0 & \cos2\psi & 0 & \sin2\psi \cr
      0 & 0 & 1 & 0 \cr
      0 & -\sin2\psi & 0 & \cos2\psi \cr
} ,  
\end{equation}
\begin{equation}
 R_3 \left( \gamma \right) := \pmatrix{
      1 & 0 & 0 & 0 \cr
      0 & \cos\gamma & -\sin\gamma & 0 \cr
      0 & \sin\gamma & \cos\gamma & 0  \cr
      0 & 0 & 0 & 1 \cr
} , 
\end{equation}
\begin{equation}
 \tanh\frac{ \alpha}{2} := \frac{ \sqrt{40\sqrt{5}-75}}{5},
 \quad \tan2\psi := 2.
\end{equation}
The compact hypersurfaces $H^{3} / \Gamma$ are constructed 
by the identifications;  
\begin{equation}
X^a\qquad \stackrel{{\small \mbox{identify}}}
{\longleftrightarrow}\qquad 
X'\,^b = (T_i)^a\,_b X^b.
\end{equation}
\begin{figure}
  \begin{center}
{\BoxedEPSF{id.eps scaled 340} }
\end{center}
 \caption[dammy]{Gray parts of surfaces of dodecahedra are 
dihedral pieces with dihedral angle $2\pi /5$ . 
The front surfaces of a hyperbolic dodecahedron are marked 
with the bold-faced letters {\bf A} $\sim$ {\bf L}. 
The surfaces marked with the letters A $\sim$ L are behind. 
The dark gray triangular parts of the front surfaces are 
identified to the light gray triangular parts of the 
opposite side. 
Five dihedral pieces consistently meet at the identified edge. }
\protect 
\end{figure}

The value of the boost angle $\alpha$ is determined 
such that the $H^{3} / \Gamma$ is a regular 
compact manifold. 
Each of the $T_{i}$ transforms 
each surface of a dodecahedron 
to an opposite side after rotating by ${3}\pi /5$. 
The rotations are necessary so that 
five dodecahedra with dihedral angle $2\pi /5$ consistently 
meet at the identified edge and add up to $2\pi$ 
as depicted in Figure 7. 

\section{Ergodic theory}

A triplet $(M, \mu, \phi_t)$ is called {\it an abstract dynamical system}, 
where $(M,\mu)$ is a measure space and 
$\phi_t :M \rightarrow M$ is a one-parameter 
group of transformations which preserve measure $\mu$.

Let $f$ be a function on $M$. 
{\it The time-average} $f^*(x)$ of $f$ 
at $x\in M$ is defined as 
$$ 
f^*(x) := \lim_{T \to +\infty} \frac{1}{T} \int_0^T
f(\phi_t(x))dt,
$$
which exists for almost every $x$~\cite{AA}.
{\it The space-average} $\bar f$ is defined as 
$$ 
\bar{f}:= \frac1{\mu(M)} \int_M f(x)d\mu.
$$

A dynamical system $(M,\mu,\phi_t)$ is {\it ergodic} 
if almost everywhere 
$f^*(x) = \bar f $, for any $f$ which is integrable with respect to 
$\mu$ (i.e. $f \in L_1(M, \mu)$). 

Let us derive a geometrical property implied by ergodicity.
Let $(M,\mu,\phi_t)$ be an ergodic abstract dynamical system on a 
compact, connected Riemannian manifold $M$.
Let $A$ be an open subset of $M$.
Define a function $f_A$ as
\begin{eqnarray*}
  f_A(x) := \left\{
    \begin{array}{ll}
      1, & \quad x\in A\\
      0, & \quad x\notin A.
    \end{array}
  \right.
\end{eqnarray*}
Then, there exists a time-average 
\begin{equation}
  f_A^*(x) = \lim_{T \rightarrow \infty}\frac{I_A(T)}{T}
  \label{eq:f_A^*}
\end{equation}
for almost every $x$, 
where $I_A(T)$ is a total length of $t$ of the intersections of 
$\{\phi_t(x) \,|\, 0\le t\le T\}$ and $A$.
The space-average is given by
\begin{equation}
  \bar{f_A} = \frac{\mu(A)}{\mu(M)}.
  \label{eq:barf_A}
\end{equation}
The ergodicity implies $f_A^*(x)= \bar{f_A}$ for any $A$ and almost
every $x$. 
From these observations, we have the following lemma.
\begin{quote}
  {\bf Lemma.}
  \\ 
  {\it Let $(M,\mu,\phi_t)$ be an ergodic dynamical system.
    For any $T_0>0$, and for almost every $x$ 
    and any open neighborhood $A$ of $x$, there exist
    $T>T_0$ such that $\phi_T(x)\in A$.}
\end{quote}

\noindent{\em Proof.}\\
Suppose there does not exist $T>T_0$ such that $\phi_T(x)\in A$, for 
almost every $x$.
It follows from (\ref{eq:f_A^*}) that $f_A^*(x)=0$ for
almost every $x$.
On the other hand, (\ref{eq:barf_A}) implies that 
$\bar{f_A}>0$.  
Thus we have $f_A^*\ne \bar{f_A}$ for almost every $x$, 
which contradicts the ergodicity. \hfill$\Box$

Let us consider a geodesic flow on a Riemannian manifold.
Let $(M,g)$ be a compact, connected Riemannian manifold, and $T_1M$ be 
its unit tangent bundle.  There is a natural Riemannian metric $\widehat
g$ on $T_1M$ induced by $g$ and a natural measure $\mu$ on $T_1M$
induced by $\widehat g$.
Consider a geodesic $\lambda_x$ on $M$
parametrized by length $\tau$ and, 
which determined by $x=(p,v)\in T_1M$ by the condition
$\lambda_x(0)=p\in M, 
\dot\lambda_x(0)=v\in T_{1p}M$.
Each $\lambda_x$ has a unique lift on $T_1M$.
Considering all geodesics we can define a 
{\em geodesic flow}
$\phi_\tau$ on $T_1M$ by 
$\phi_\tau(x):=(\lambda_x(\tau),\dot\lambda_x(\tau))$.

We essentially used the following theorem in the proof of the main
proposition of this paper.
\begin{quote}
  {\bf Theorem.}\\
  (Lobatchewsky--Hadamard--Anosov~\cite{AA}) \\
  {\it Let $M$ be a compact, connected Riemannian manifold
    with negative curvature.
    $T_{1}M$ be a unit tangent bundle of $M$.  
   Then, the geodesic flow on $T_1M$ is ergodic.}
\end{quote}
From the Theorem and the Lemma above, we have the following, which
is actually used in our proof.
\begin{quote}
  {\bf Proposition.}\\
  {\em Let $M$ be a compact, connected Riemannian manifold
  with negative curvature.  
  Let $\lambda_x$ be a geodesic defined above.
  For any $T_0>0$, and for almost every $x=(p,v)$
  and any open neighborhood $O$ of $p$ in $M$, 
  there exist $T>T_0$ such that $\lambda_x(T)\in O$ and 
  the (unit) tangent vector $\dot\lambda_x(T)$ is arbitrarily close to 
  $\dot\lambda_x(0)$. }
\end{quote}
 
\end{document}